\documentclass[12pt,onecolumn,draftcls]{IEEEtran}
\usepackage{amsmath,epsfig}
\newcommand{\y}{\mathbf{y}}

\newcommand{\Y}{\mathbf{Y}}

\newcommand{\Hnull}{\mathcal{H}_0}
\newcommand{\Halt}{\mathcal{H}_1}

\usepackage{pifont}
\usepackage{graphicx}
\usepackage{amssymb}
\usepackage{amsmath}
\usepackage{cancel}
\usepackage[normalem]{ulem}
\usepackage[dvips]{color}
\usepackage{algorithm}  
\newfont{\fsc}{eusm10}                         
\DeclareMathOperator*{\argmin}{arg\,min}
\DeclareMathOperator*{\argmax}{arg\,max}

\usepackage{lineno}
\modulolinenumbers[1]

\usepackage{algorithmic}
\usepackage{algorithm}

\title{Blind Spectrum Sensing in Cognitive Radio over Fading Channels and Frequency Offsets}
\author{
\IEEEauthorblockN{Ido Nevat$^1$,
                  Gareth W. Peters $^{2,3}$,
                  and Jinhong Yuan$^4$\\
\IEEEauthorblockA{$^1$
Wireless \& Networking Tech. Lab, CSIRO, Sydney, Australia.}\\
\IEEEauthorblockA{$^2$
School of Mathematics and Statistics, University of NSW, Sydney, Australia.}\\
\IEEEauthorblockA{$^3$
CSIRO Mathematical and Information Sciences, Sydney, Australia.}\\
\IEEEauthorblockA{$^4$
School of Electrical Engineering, University of NSW, Sydney, Australia.}} }
\begin{document}
\maketitle
\begin{abstract}
This paper deals with the challenging problem of spectrum sensing in cognitive radio. 
We consider a stochastic system model where the Primary User (PU) transmits a periodic signal over fading channels. 
The effect of frequency offsets due to oscillator mismatch, and Doppler offset is studied. We show that for this case the Likelihood Ratio Test (LRT) cannot be evaluated pointwise. We present a novel approach to approximate the marginilisation of the frequency offset using a single point estimate. This is obtained via a low complexity Constrained Adaptive Notch Filter (CANF) to estimate the frequency offset.
Performance is evaluated via numerical simulations and it is shown that the proposed spectrum sensing scheme can achieve the same performance as the ``near-optimal" scheme, that is based on a bank of matched filters, using only a fraction of the complexity required.
\end{abstract}
\vspace{-0.5cm}
\vspace{-0.5cm}
\section{Introduction}
In recent years, cognitive radio has attracted intensive research focus due to the pressing demand for efficient usage of the frequency spectrum \cite{haykin2005cognitive}. In a cognitive radio system, secondary radio users try to find "blank spaces", in which the licensed
frequency band is not being used by primary radio users, for communications. A key problem in cognitive radio is that the
secondary users need to vacate the frequency band as quickly as possible if the corresponding primary radio emerges, and begins transmission.

Spectrum sensing is a mandatory functionality in any CR-based wireless system that shares spectrum bands with primary services, such as the IEEE
802.22 standard \cite{IEEE802_22}, which proposes to reuse vacant spectrum in the TV broadcast bands.
There has been significant research on spectrum sensing for cognitive radio, see \cite{ghasemi2008spectrum}, \cite{akyildiz2008survey} for an overview.
Essentially, spectrum sensing is a decision making or classification problem, of the form involving first an estimation stage, followed a decision stage. The secondary network needs to make a decision between two possible hypotheses given an observation vector: that the frequency band is either occupied or vacant.
The more knowledge we have on the nature of the primary user's signal, the more reliable our decision process will become. If no prior knowledge of the primary user's signal, the energy detector based approaches (also called radiometry) are the most common  for spectrum sensing, mainly due to their low computational complexity, see \cite{yucek2009survey} and references within. If additional prior knowledge about the primary signal exists, more advanced techniques can be explored. For example, if the primary signal is \textit{a priori} known, the matched filter is optimal in the sense that it maximises the Signal to Noise Ratio (SNR) \cite{yucek2009survey}, \cite{vantrees1968dea}. 

In practical systems, such \textit{a priori} knowledge often exists in the form of a pilot signal, which is used by the primary network, enabling a Waveform-Based Sensing \cite{yucek2009survey}. For example, in the standard of digital TV (DTV) system, made by advanced television standard committee (ATSC), there are multiple sinusoid pilots located at different frequencies \cite{IEEE802_22}. 

\textbf{Previous works:} A few papers considered the problem of spectrum sensing using a pilot signal in the form of sinusoids. These include \cite{li2008quickest}, \cite{zheng2010spectrum} and \cite{li2009adaptive} where the detection scheme was designed for the case that the primary user transmits at a \textbf{known deterministic frequency} and the amplitude/channel may or may not be known \textit{a priori}.\\
\textbf{Contribution:} in contrast to those papers, we extend the system model and consider two practical effects which are of significance in a wireless communication system: first, we consider the case of \textbf{unknown Rayleigh fading channel}. Second, we allow for \textbf{frequency offsets} due to Doppler offset and mismatched oscillators being present in the communications system. These two assumptions make the sensing algorithm complex. 
In order to overcome this difficulty, we shall present two ``close to optimal" sensing algorithms that are based on a bank of matched filters, and on a frequency estimation (periodogram) approaches. While these solutions perform close to optimal, their high complexity may prevent them from being practical. Instead, we propose a low complexity algorithm that is based on the principle of Generalised Likelihood Ratio Test (GLRT), where we condition the LRT on an estimate of the nuisance frequency parameter (the estimate is not the Maximum Likelihood (ML) estimate, and therefore, the proposed algorithm is not the GLRT). The estimation of the nuisance frequency parameter is based on the Constrained Adaptive Notch Filter (CANF). Then, we perform a single matched filter centered on the estimated frequency, providing very good performance with only a fraction of the computational complexity.\\
The following notation is used throughout: random variables are denoted by upper case letters and their realizations by lower case letters.
In addition, bold will be used to denote a vector or matrix quantity.
\vspace{-0.5cm}
\section{System Model}\label{System_Model}
\begin{figure}
    \centering
        \epsfysize=7cm
        \epsfxsize=11cm
        \epsffile{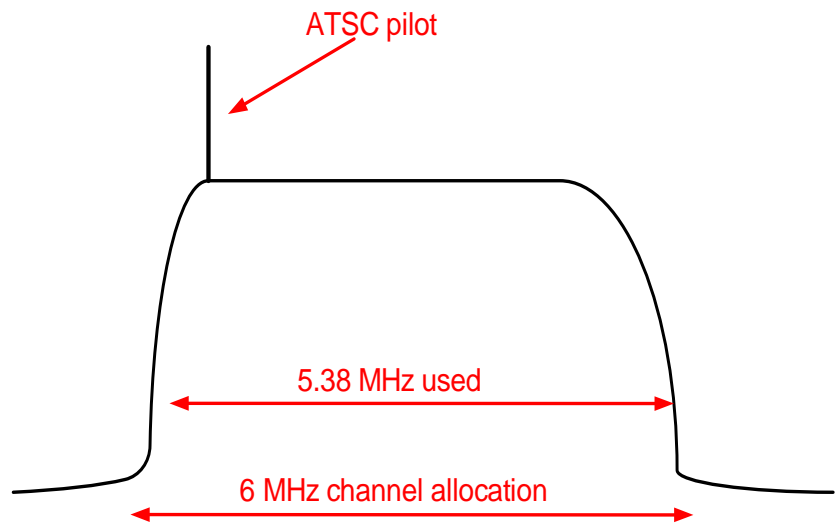}
        \caption{Spectrum of ATSC channel spectrum. The channel occupies $6$ MHz and is relatively flat except for the pilot signal located in $310$ kHz above the lower edge
of the channel.  }
    \label{fig:HDTV}
\end{figure}

Consider a secondary radio communication network using a single sensor. 
The secondary sensor needs to monitor the activity of the primary network. 
We assume that the primary radio signal uses a pilot signal in the form of a sinusoid, as in \cite{li2008quickest}, \cite{zheng2010spectrum} and \cite{li2009adaptive}, see Fig \ref{fig:HDTV}.
For example, in High-definition TV (HDTV) systems, developed by advanced television standard committee (ATSC), there are multiple sinusoid pilots located at different frequency points.\\
\subsection{Model Assumptions}
We introduce the system model for spectrum sensing
\begin{itemize}
\item \textit{Assume a secondary radio communication network using a single receive antenna.}
\item \textsl{The primary is \textbf{active} in a frame  (block) of length $M$ samples, with probability $P\left(\Hnull\right)$ and \textbf{idle} with probability $P\left(\Halt\right)$.} 
\item \textit{The observation at the sensor can be written as the following binary hypothesis:
\begin{align}
\label{binary_hypothesis}
\begin {cases}
\begin{split}
    &\Hnull: Y(m)=V(m),\; m=1,\ldots,M\\
    &\Halt: Y(m)=H \sin\left(m \Omega + \Theta\right)+V(m),\;m=1,\ldots,M,
\end{split}
\end {cases}
\end{align}
where $H$ is the unknown random channel gain (assumed to be fixed throughout a frame), $\Omega$ is the fixed unknown random angular frequency of the primary user signal, which is not known \textit{a-priori}, due to oscillator mismatch and Doppler offset, and $\Theta$ is the random unknown phase offset of the pilot signal, and  $V(m)$ is the random additive noise.}
\end{itemize}
\subsection{Prior specification}
We present the relevant aspects of the Bayesian model.
\begin{itemize}
\item \textit{The channel gain $H$ is Rayleigh fading. Therefore its density function is
\begin{align}
p\left(h\right) =
\begin {cases}
\begin{split}
     &\frac{h}{\sigma^2_h}\exp\left(-\frac{h^2}{2 \sigma^2_h}\right), \; h\geq 0 \\
    &0 \;\;\;\;\;\;\;\;\;\;\;\;\;\;\;\;\;\;\;\;\;\;\;\;,\text{otherwise},
\end{split}
\end {cases}
\end{align}
where $\sigma^2_h$ is the known variance of the channel gain.
\item The phase offset $\Theta$ is random and follows a uniform density function
$\Theta \sim U\left[0,2 \pi\right).$
}
\item \textit{The angular frequency of the primary user's signal, $\Omega$, can be parametrised by density function distribution, $p\left(\omega\right)$, over the support $\left[\overline{\omega} - \epsilon, \overline{\omega} + \epsilon\right]$, with $\overline{\omega}$ being the nominal known angular frequency and $\epsilon$ as the maximal offset, determined by the Doppler offset and oscillators mismatch. In this paper, for ease of presentation, we assume a uniform prior. 
}
\item{\textit{The received signal is corrupt by zero-mean i.i.d. additive white Gaussian noise (AWGN) $V(m) \sim \mbox{CN}\left(0,\sigma_v^2\right)$, with a known variance $\sigma_v^2$ and power spectral density (PSD) $\text{N}_0$. 
}}
\end{itemize}
\section{Spectrum Sensing - Problem Definition}
The objective of spectrum sensing is to make a decision on the binary hypothesis testing (choose $\Hnull$ or $\Halt$) based on the received signal.
\subsection{Decision criterion}
Here we formulate the problem using Bayes' criterion \cite{vantrees1968dea}.
In doing so, two assumptions are made. First, the probabilities $P\left(\Hnull\right)$ and $P\left(\Halt\right)$ are known.
The second assumption is that a cost, $C_{xy}$, is assigned to each possible decision. $C_{xy}$ is the associated cost of making a decision $\mathcal{H}_x$,  given that the true hypothesis is $\mathcal{H}_y$.

The problem of designing the decision rule can be modeled as an optimization problem whose objective is to minimize the cost function

\begin{align}
\begin{split}
C &=
P\left(\Hnull\right)
\left(C_{00}\int_{A_0}p\left(\y_{1:M}|\Hnull\right)d\y_{1:M}
+C_{10}\int_{A_1}p\left(\y_{1:M}|\Hnull\right)d\y_{1:M}\right)\\
&+P\left(\Halt\right)
\left(C_{01}\int_{A_0}p\left(\y_{1:M}|\Halt\right)d\y_{1:M}
+C_{11}\int_{A_1}p\left(\y_{1:M}|\Halt\right)d\y_{1:M}\right).
\end{split}
\end{align}
It can be shown that the optimum decision rule is a likelihood-ratio test given by
\begin{equation}
\label{LRT}
\Lambda\left(\Y_{1:M}\right) \triangleq
\frac{p\left(\y_{1:M}|\Halt\right)}{p\left(\y_{1:M}|\Hnull\right) }
\begin{array}{c}
\stackrel{\Halt}{\geq} \\
\stackrel{<}{\Hnull}
\end{array}%
\frac{P\left(\Hnull\right)}{P\left(\Halt\right)}
\frac{C_{10}-C_{00}}{C_{01}-C_{11}}\triangleq \gamma,
\end{equation}
where $C_{xy}$ is the associated cost of making a decision $\mathcal{H}_x$, given that the true hypothesis is $\mathcal{H}_y$, and we define $\y_{1:M} \triangleq \left[\y\left(1\right),\ldots, \y\left(M\right)\right]$.\\
The major difficulty in using the LRT is its requirement on obtaining the exact distributions under each hypothesis in (\ref{binary_hypothesis}).
Under the NULL hypothesis, the observations are independent, and the evaluation of the evidence, $p\left(\y_{1:M}|\Hnull\right)$, can be decomposed as
\begin{equation}
p\left(\y_{1:M}|\Hnull\right) = \prod_{m=1}^M p\left(y(m)|\Hnull\right)
=\frac{1}{\sqrt{2 \pi \sigma_v^2}}\prod_{m=1}^M \exp\left(-\frac{1}{2 \sigma_v^2}y^2(m)\right).
\end{equation}
The distribution of the alternative, $p\left(\y_{1:M}|\Halt\right)$, may be harder to obtain depending on the knowledge of the system parameters.
Here we develop the solution for several cases with different levels of knowledge of the system parameters.
\vspace{-0.2cm}
\subsection {Case I: no frequency offset ($\omega= \overline{\omega}$) and known channel gain $H$}
With no frequency offset present ($\omega$ is known exactly) and the channel gain $h$ is known \textit{a-priori}, the likelihood ratio has the following expression \cite{vantrees1968dea}
\begin{equation}
\label{case_1}
\Lambda^I\left(\Y_{1:M}\right) = \frac{p\left(\y_{1:M}|\Halt\right)}{p\left(\y_{1:M}|\Hnull\right)}=
\exp\left(\frac{- M h^2}{2 \sigma^2_w} \right) I_0\left(\frac{2h}{\sigma^2_w} r\right),
\end{equation}
where $I_0$ is the modified Bessel function and $r$ is defined as
\begin{subequations}
\begin{align}
\label{MF_def}
r\left(\omega\right) &\triangleq \sqrt{ \left(y_c^2+y_s^2\right)}, \;\;\text{where}\\
y_c\left(\omega\right) &\triangleq \sum_{m=1}^M y(m) \cos\left(m \omega\right),\\
y_s\left(\omega\right) &\triangleq \sum_{m=1}^M y(m) \sin\left(m \omega\right).
\end{align}
\end{subequations}
Here a single matched filter is required to perform the LRT.
\vspace{-0.2cm}
\subsection {Case II: no frequency offset ($\omega= \overline{\omega}$) and unknown channel gain $H$}
With no frequency offset present, but the realisation of the channel gain $h$ unknown \textit{a-priori}, the LRT can be evaluated exactly by marginalising (\ref{case_1}) over the unknown channel,$H$, as
\begin{equation}
\label{L_unknown_channel}
\begin{split}
\Lambda^{II}\left(\Y_{1:M}\right) &=
\int^{\infty}_{0}\Lambda^{I}\left(\Y_{1:M}|h\right)p\left(h\right)d h \\
&=\int^{\infty}_{0}
\exp\left(\frac{- M h^2}{2 \sigma^2_w} \right) I_0\left(\frac{2h}{\sigma^2_w} r\right)
\frac{h}{\sigma^2_h}\exp\left(-\frac{h^2}{2 \sigma^2_h}\right)
d h \\
&=\frac{\sigma^2_w}{\sigma^2_w+M \sigma^2_a}
\exp\left(\frac{ 2\sigma^2_a}{\sigma^2_w\left(\sigma^2_w+ M\sigma^2_a\right)} r^2\left(\omega\right)\right),
\end{split}
\end{equation}
where $r\left(\omega\right)$ is defined in (\ref{MF_def}).
Again, a single matched filter is required to perform the LRT.
\vspace{-0.2cm}
\subsection {Case III: unknown frequency and unknown channel gain}\label{unknown_frequency}
When a frequency offset is present and the channel gain $h$ is unknown \textit{a-priori}, we consider (\ref{L_unknown_channel}) and marginalise over the unknown random frequency $\omega$,
\begin{align}
\begin{split}
\label{LR_blind}
\Lambda^{III}\left(\Y_{1:M}\right) &=
\int^{\overline{\omega} + \epsilon}_{\overline{\omega} - \epsilon}\Lambda^{II}\left(\Y_{1:M}|\omega\right)p\left(\omega\right)d \omega \\
&=\frac{\sigma^2_w}{\sigma^2_w+M \sigma^2_a}
\int^{\overline{\omega} + \epsilon}_{\overline{\omega} - \epsilon}
\exp\left(\frac{ 2\sigma^2_a}{\sigma^2_w\left(\sigma^2_w+ M\sigma^2_a\right)} r^2\left(\omega\right)\right)
p\left(\omega\right)d \omega.
\end{split}
\end{align}
The integral in (\ref{LR_blind}) is not analytic and requires approximation techniques, and we shall derive numerical approximations of (\ref{LR_blind}) in the following sections. 
\vspace{-0.2cm}
\section{High-Complexity Blind Spectrum Sensing Estimation}
In this section we briefly present two possible approximations which have high complexity relative to the solution we propose in this paper. They act as benchmarks for comparison with our solution. The first approach is based on a bank of matched filters. We approximate $p\left(\omega\right)$ using a discrete density function $p_d\left(\omega\right)$ with $K$ discrete values as
\begin{align}
p_d\left(\omega\right) \approx \sum_{k=1}^K p\left(\omega_k\right) \delta\left(\omega - \omega_k\right),
\end{align}
where 
$
\omega_k = \left(\overline{\omega} - \epsilon\right)+k \Delta ,\; k=\left\{1,2,\ldots,K\right\},\;\;
K \triangleq \frac{2 \epsilon  }{\Delta }.$\\
We can now approximate (\ref{LR_blind}) using a discretization, written as
\begin{equation}
\label{LR_blind_approximate}
\Lambda\left(\Y_{1:M}\right) \approx \sum_{k=1}^K \Lambda\left(\Y_{1:M}|\omega\right) p\left(\omega_k\right)
=\frac{\sigma^2_w}{\sigma^2_w+M \sigma^2_a}
\sum_{k=1}^K p\left(\omega_k\right) \exp\left(\frac{ 2\sigma^2_a}{\sigma^2_w\left(\sigma^2_w+ M\sigma^2_a\right)} r^2\left(\omega_k\right)\right).
\end{equation}
Here, the LR function is evaluated by using a bank of $K$ matched filters, one per frequency $\omega_k$. This solution has high complexity since it performs a weighted average of (\ref{LR_blind}). Clearly, with this approach a strong trade-off between performance and complexity burden occurs . The matched filter is sensitive to frequency mismatch (see Section \ref{COMPUTATIONAL_COMPLEXITY_AND_ANALYSIS} for analysis) and it is therefore desirable to set $K$ to be very large. This would make the gap between consecutive discrete frequencies, $\Delta \omega$, very small and make the frequency mismatch ($\Omega$ vs. $\omega_k$) negligible. This however would result in a very costly implementation. In cases where only a few matched filters are used, it is likely that a frequency mismatch will occur, leading to poor performance.

The second approach is based on a GLRT and discrete Fourier transform. This solution first produces a periodogram, followed by a LRT conditioned on the maximum value of the frequency obtained from the periodogram. This has high complexity due to the construction of the periodogram.
In the GLRT, we condition on frequency $\Omega$ obtained from a point estimate from the periodogram \cite{vantrees1968dea}. Hence, we approximate (\ref{LR_blind}) as
\begin{equation}
\label{ANF_MF_blind}
\widehat{\Lambda}^{III}\left(\Y_{1:M}\right) 
\approx
\frac{p\left(\y_{1:M}|\widehat{\Omega}_{\text{ML}},\Halt\right)}{p\left(\y_{1:M}|\Hnull\right)}
=
\frac{\sigma^2_w}{\sigma^2_w+M \sigma^2_a}
\exp\left(\frac{ 2\sigma^2_a}{\sigma^2_w\left(\sigma^2_w+ M\sigma^2_a\right)} r^2\left(\widehat{\Omega}_{\text{ML}}\right)\right).
\end{equation}
It is well known that $\widehat{\Omega}_{\text{ML}}$ can be asymptotically obtained by maximising the periodogram \cite{vantrees1968dea}, so that 
$\widehat{\Omega}_{\text{ML}} = \argmax_{\omega}\left|\sum_{m=1}^M	y\left(m\right)	\exp^{-j \omega m}\right|,$ where $j \triangleq \sqrt{-1}$.
The accuracy of the frequency estimator depends on the number of samples in the frame, $M$. 
As with the bank of matched filters, a fine grid of frequencies is required, resulting in a highly computational algorithm. If, on the other hand, we used only a coarse grid with only a few frequencies, that would result in a significant loss of accuracy and high estimation error.
\section{Novel Low Complexity Blind Spectrum Sensing Estimation}
In this section we present a novel algorithm to perform a low complexity spectrum sensing for Case III. As in the case of approach two, our solution is also based on the GLRT, but replaces the grid search required by the periodogram construction with an adaptive notch filter based frequency estimator. This reduces the computational complexity significantly for the same estimation accuracy.
Here we develop a non-standard solution for the GLRT which involves designing a notch filter which performs adaptive frequency estimation. The key to our solution is to utilise the result involving the representation of the transmitted signal under $\Halt$ in (\ref{binary_hypothesis}) as a $2$-nd order autoregressive process, obtained via a trigonometric identity,
\begin{align}
\label{trigonometric_identity}
\sin\left(m \Omega + \Theta\right) = 2 \cos\left(\Omega\right) \sin\left((m-1) \Omega + \Theta\right)+ \sin\left((m-2) \Omega + \Theta\right).
\end{align}
In the frequency domain, the transmitted signal is represented by Dirac masses, with unknown locations. Using (\ref{trigonometric_identity}), we can estimate this location via a localised filter in the family of notch filters. A notch filter is a filter that contains a null in its frequency response characteristics. 
Here, for simplicity, we concentrate on a 2-nd order Infinite Impulse Response (IIR) which contains a pair of complex-conjugate zeros on the unit circle and a pair of complex-conjugate poles at the same frequency inside the unit circle, and has the following transfer function
\begin{equation}
\label{TransferFunciton}
H(z) = \frac{1- \beta(m)  z^{-1}+z^{-2}}{1-\rho(m) \beta(m) z^{-1}+\rho^2(m) z^{-2}},
\end{equation}
where the values $\beta(m)$ determine the centre of the notch filter frequency, and $0<\rho(m) < 1$ defines the location of the poles inside the unit disk. 
This design has the properties of having a symmetric frequency response and a narrow bandwidth, provided that $\rho(m)$ is close to $1$.
This filter is simple to design, requiring the estimation of two parameters, keeping complexity low, whilst providing narrow-band frequency selectivity. We now present the specific details of the proposed frequency estimation algorithm.\\
The output of the filter, $s(m)$, as defined by (\ref{TransferFunciton}) can be expressed as
\begin{equation}
\label{ANF}
s(m) = y(m)+\beta(m) y(m-1) + y(m-2)
 -\rho(m) \beta(m) s(m-1)-\rho^2(m) s(m-2).
\end{equation} 
We formulate the joint optimisation for $\beta$ and $\rho$ using the following criterion
\begin{equation}
\label{cost_function}
\begin{split}
\left(\widehat{\beta}, \widehat{\rho}\right) =  \argmin_{
\begin{array}{l}
\beta_{\min} \leq \beta \leq \beta_{\max},\\
\rho \leq \rho_{\max} 
\end{array}
}
\frac{1}{M}\sum_{m=1}^M \left(s^2(m)+\frac{1}{\rho(m)}\right),
\end{split}
\end{equation} 
where $\beta_{\min}, \beta_{\max}$ are determined by the allowed frequency offset $\epsilon$,
and $\rho_{\max} \leq 1$ to ensure the filter is stable, according to Lyapunov stability criterion \cite{lyapunov1994general}.
This cost function is designed to achieve two goals: the first is to minimise the variance at the output of the filter (this is given by the first term in the summation on the RHD term); and penalising for poles located far from the unit disk, as this makes the bandwidth of filter wider and also creates bias in the resulting estimated frequency (this is given by the second term in the summation on the RHD term).
The bandwidth of the notch filter is determined by $\rho(m)$ as $\text{BW} = \pi \left(1-\rho(m) \right)$ \cite{stoica1988performance}.
The estimated frequency, $\widehat{\Omega}$, can be retrieved by $ \widehat{\Omega}= \arccos \left(-\frac{\beta(M)}{2}\right),$ and used to approximate the GLRT in (\ref{ANF_MF_blind}). Although this estimator is asymptotically biased, the bias can be made arbitrarily small by choosing $\rho \rightarrow 1$ \cite{stoica1988performance}.
However, the initial radius, $\rho(1)$, should be set such that $\text{BW} \geq 2 \epsilon$, to ensure that the realised frequency lies within the filter's range.

Direct optimisation of (\ref{cost_function}) is difficult due to its non-linearity. However, this problem can be easily solved in a sequential manner, by utilising adaptive filter theory. Here we use the steepest descent approach to minimise the associated cost function in (\ref{cost_function}). This results in adapting $\beta$ and $\rho$ and tracking the location of the null frequency as follows
\small
\begin{subequations}
\begin{align}
\label{adapting_beta}
\beta(m) &= \beta(m-1) -\mu_{\beta} \frac{\partial s^2(m)}{\partial \beta}
=\beta(m-1) -2 \mu_{\beta}  s(m)\left( y(m-1) -\rho s(m-1)\right),\\
\label{adapting_rho}
\rho(m) &= \rho(m-1)   -\mu_{\rho} \frac{\partial}{\partial \rho}\left( s^2(m)+\frac{1}{\rho(m)}\right)=
 \rho(m-1)  +2 \mu_{\rho} s(m) \left(\beta(m) s(m-1) +2\rho(m) s(m-2)\right)+\frac{\mu_{\rho}}{\rho^{2}(m)},
\end {align}
\end{subequations} 
\normalsize
where 
$\mu_{\beta}$ and $\mu_{\rho}$ are predefined step size parameters.
Since the support of the frequency is bounded over the interval $\left[\overline{\omega} - \epsilon, \overline{\omega} + \epsilon\right]$, the values of $\beta(m)$ can be constrained by
$	-2 \cos \left(\overline{\omega} - \epsilon\right) \geq \beta(m) > 
-2 \cos \left(\overline{\omega} + \epsilon\right)$.
These constraints are useful in case of low SNR and prevent the filter from wondering outside the boundaries of the allowed frequencies.
For the first sample, $\left(m=1\right)$, we center the filter frequency at $\beta(1) = -2 \cos \left(\overline{\omega}\right)$ and set $\rho(1)=
1-\frac{2 \epsilon}{\pi}$.
\vspace{-0.2cm}
\section{Computational Complexity and Analysis} \label{COMPUTATIONAL_COMPLEXITY_AND_ANALYSIS}
\textbf{Computational Complexity}: here we analyse the computational complexity involved in the proposed algorithms. We define the overall complexity of an algorithm as the number of complex multiplications (\texttt{CM}) and  complex additions (\texttt{CA}) required.\\
\textbf{Energy Detector} - the test statistics is $\mathbb{T}(\Y_{1:M}) = \sum_{m=1}^M  \left|\y(m)\right|^2$. The computational complexity is therefore $M\left(\texttt{CM} + \texttt{CA}\right)$.\\
\textbf{Single Matched Filter} - the test statistics is defined in (\ref{MF_def}). Obtaining $r\left(\omega\right)$ involves evaluating $y_c\left(\omega\right)$ and $y_s\left(\omega\right)$. The total computational effort is therefore $2 M\left(\texttt{CM} + \texttt{CA}\right)$.\\
\textbf{CANF based Detector} - the steps involved in the proposed algorithm are: 
\begin{enumerate}
	\item Evaluation of the filter in (\ref{ANF}): $3 M\left(\texttt{CM} + \texttt{CA}\right)$.
	\item Adapting $\beta $ in (\ref{adapting_beta}): $2 M\left(\texttt{CM} + \texttt{CA}\right)$.	
	\item Adapting $\rho $ in (\ref{adapting_rho}): $3 M\left(\texttt{CM} + \texttt{CA}\right)$.	
\item Performing a single matched filter: $2 M\left(\texttt{CM} + \texttt{CA}\right)$.	
\end{enumerate}
 Therefore, the overall complexity of the proposed algorithm is  $10 M\left(\texttt{CM} + \texttt{CA}\right)$.\\
\textbf{Performance Analysis}: we now provide an analysis regarding the performance gain obtained by the CANF algorithm over the Energy Detector and the Matched filter.
\begin{itemize}
	\item CANF vs. Energy Detector: here we show the SNR ratio gain of the CANF over the Energy detector. We do so by evaluating their ratio of SNRs
\begin{align}
\frac{\text{SNR}_{\text{CANF}} }{\text{SNR}_{\text{ED}}} = 
\frac{P_s}{\int^{\overline{\omega} - \epsilon}_{\overline{\omega} + \epsilon} \text{N}_0 \;d \omega}
/\left(\frac{P_s}{\int_{\Phi} \text{N}_0 \;d \omega}\right)=
\frac{\int_{\Phi} \text{N}_0 \; d \omega}
{\int^{\overline{\omega} - \epsilon}_{\overline{\omega} + \epsilon} \text{N}_0 \;d \omega}
=\frac{\Phi}{2 \epsilon} >> 1
\end {align}
where $P_s$ is the signal's energy, $\text{N}_0$ is the PSD of the AWGN and $\Phi$ is the system's bandwidth. Thus, the narrower the notch filter is, the greater the SNR improvement, providing better performance. This is directly related to the second term of the optimisation formulation in (\ref{cost_function}).
\item Matched Filter analysis: here we demonstrate the sensitivity of the matched filter to frequency offsets. In the complex domain the test statistic, $r\left(\overline{\omega}\right)$ in (\ref{MF_def}) can be expressed as 
\small
\begin{align}
\begin{split}
r\left(\overline{\omega}\right)| \Halt&=
\left|\sum_{m=1}^M  Y(m) \exp\left(-j m \Omega\right)\right| =
\left|\sum_{m=1}^M  \left(H \exp\left(j m \Omega+\Theta\right) +V(m) \right) \exp\left(-j m \overline{\omega}\right)\right| \\
&
\underbrace{\approx}_{\text{zero mean AWGN}}
 \left|H\right|  \underbrace{\left|\exp\left(j \Theta\right)\right|}_{1} \left|\sum_{m=1}^M  \exp\left(j m \Omega\right) \exp\left(-j m \overline{\omega}\right)\right|
\underbrace{= }_{ \delta \triangleq \Omega -  \overline{\omega}}
\left|H\right| \left|\sum_{m=1}^M  \exp\left(j m \delta\right)\right| \underbrace{\rightarrow }_{M>> 2 \pi \delta}0 .
\end{split}
\end{align}
\normalsize
Hence, the test statistic converges to $0$ for frame lengths that are larger than the period of the frequency offset, $\delta$, rendering the matched filter incapable of performing robust LRT.
\end{itemize}
%
%
%
\section{Simulation Results}
\begin{figure}
    \centering
        \epsfysize=10cm
        \epsfxsize=12cm
        \epsffile{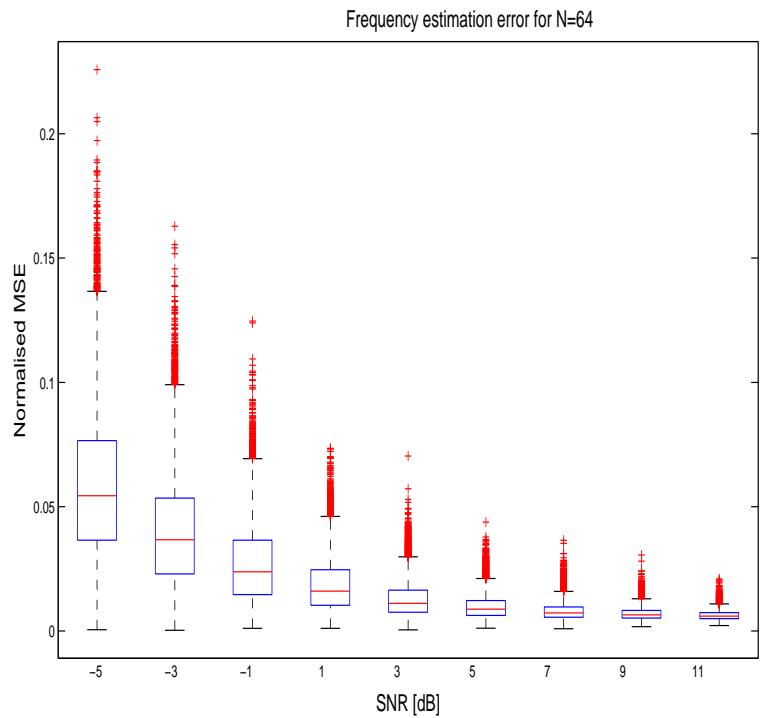}
            \caption{Frequency estimation error of the adaptive notch filter Vs. SNR for $N=64$ }
    \label{fig:estimation_error}
\end{figure}

We present the performance of the proposed algorithm and comparison via simulations.
We begin by evaluating the performance of the adaptive notch filter to perform frequency estimation.
We tested the algorithm for a frame length of $M=64$. The nominal angular frequency was set to $1.9635$ and the maximal allowed offset was  set to $0.98$. The realised angular frequency was set to $2.45$. Fig. \ref{fig:estimation_error} presents boxplot results for normalised frequency estimation error of the CANF, which clearly shows the good performance obtained by the notch filter based frequency estimator.

\begin{figure}
    \centering
        \epsfysize=9cm
        \epsfxsize=5.2cm
        \epsffile{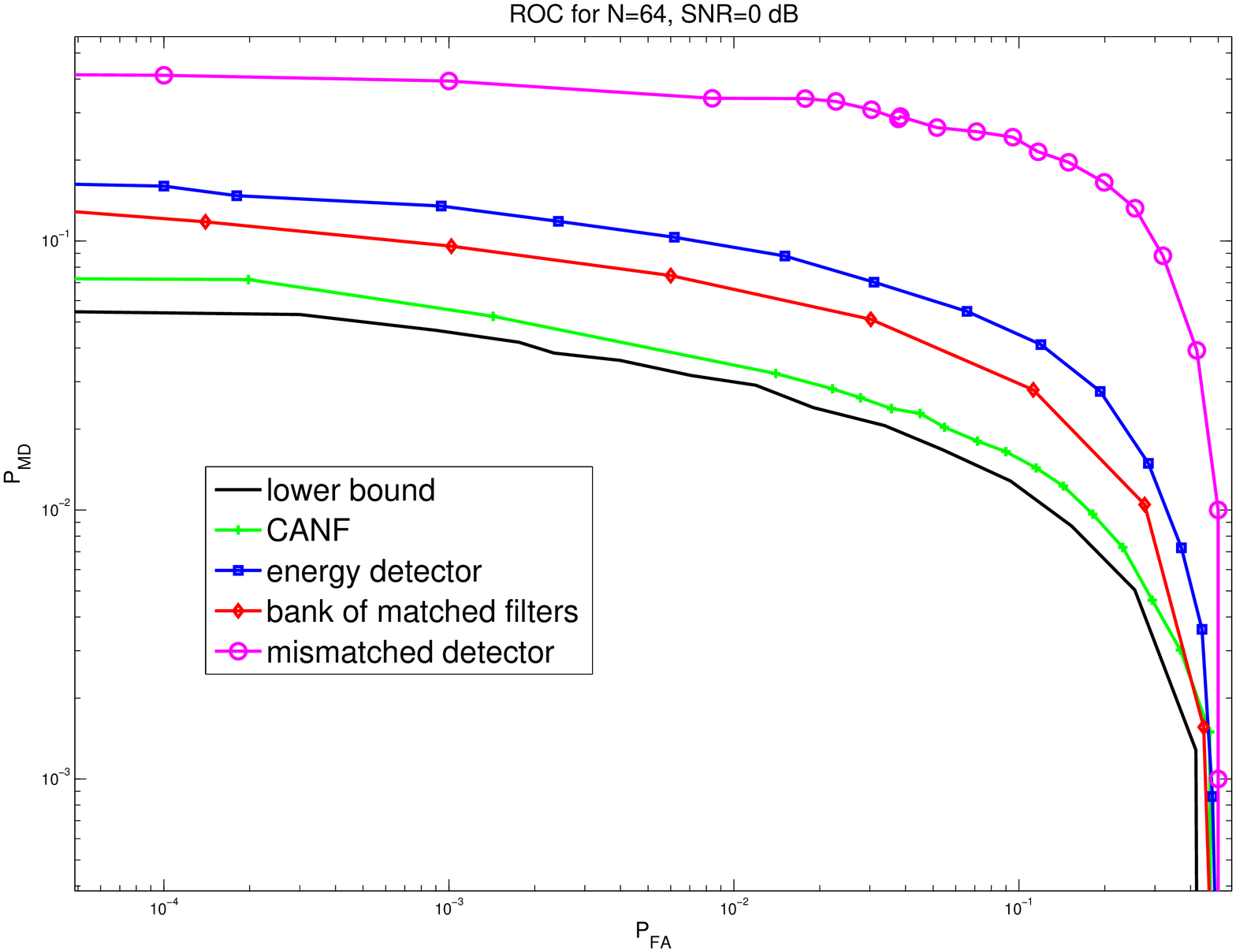}
        \epsfysize=9cm
        \epsfxsize=5.2cm
        \epsffile{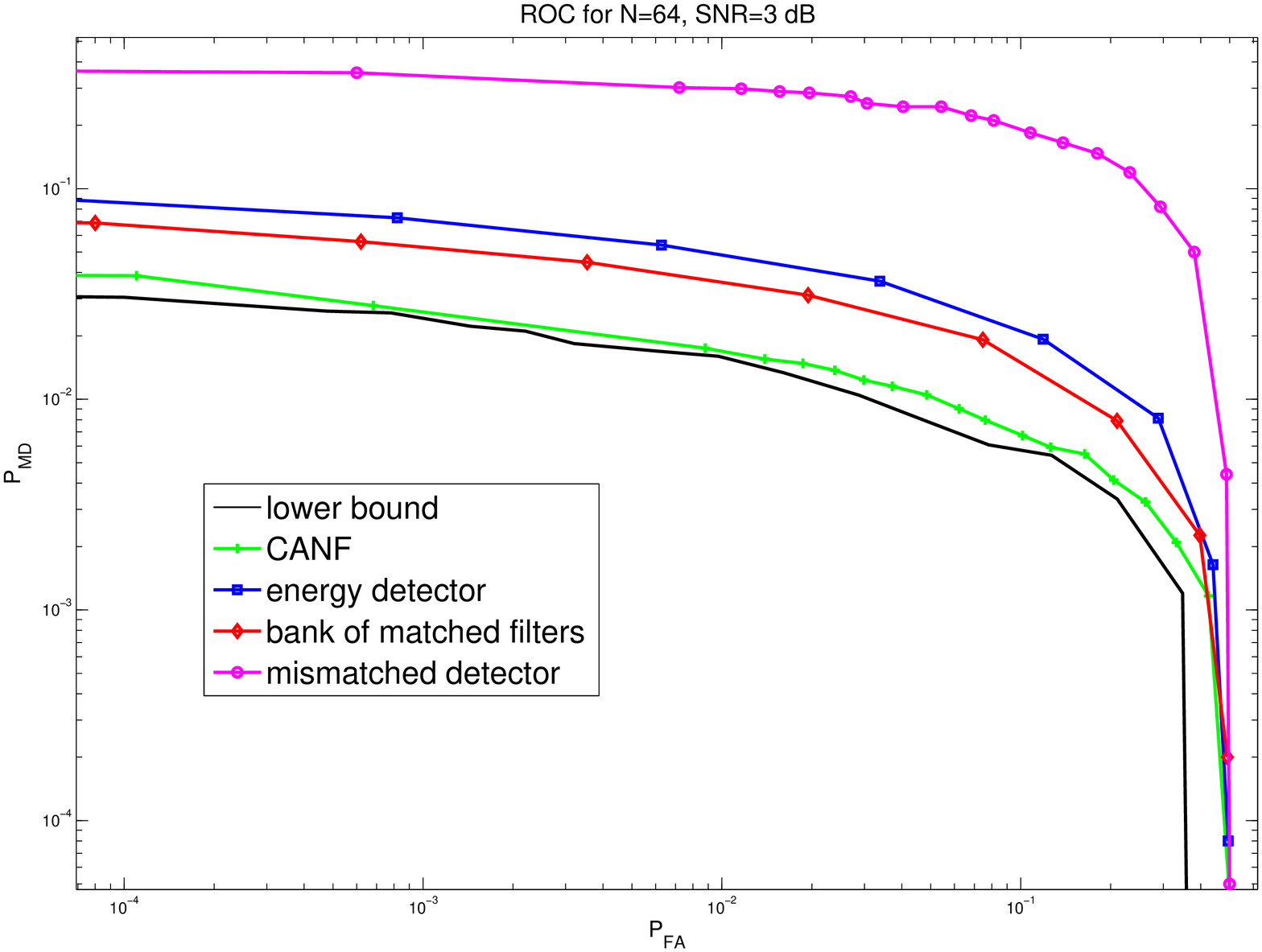}        
        \epsfysize=9cm
        \epsfxsize=5.2cm
        \epsffile{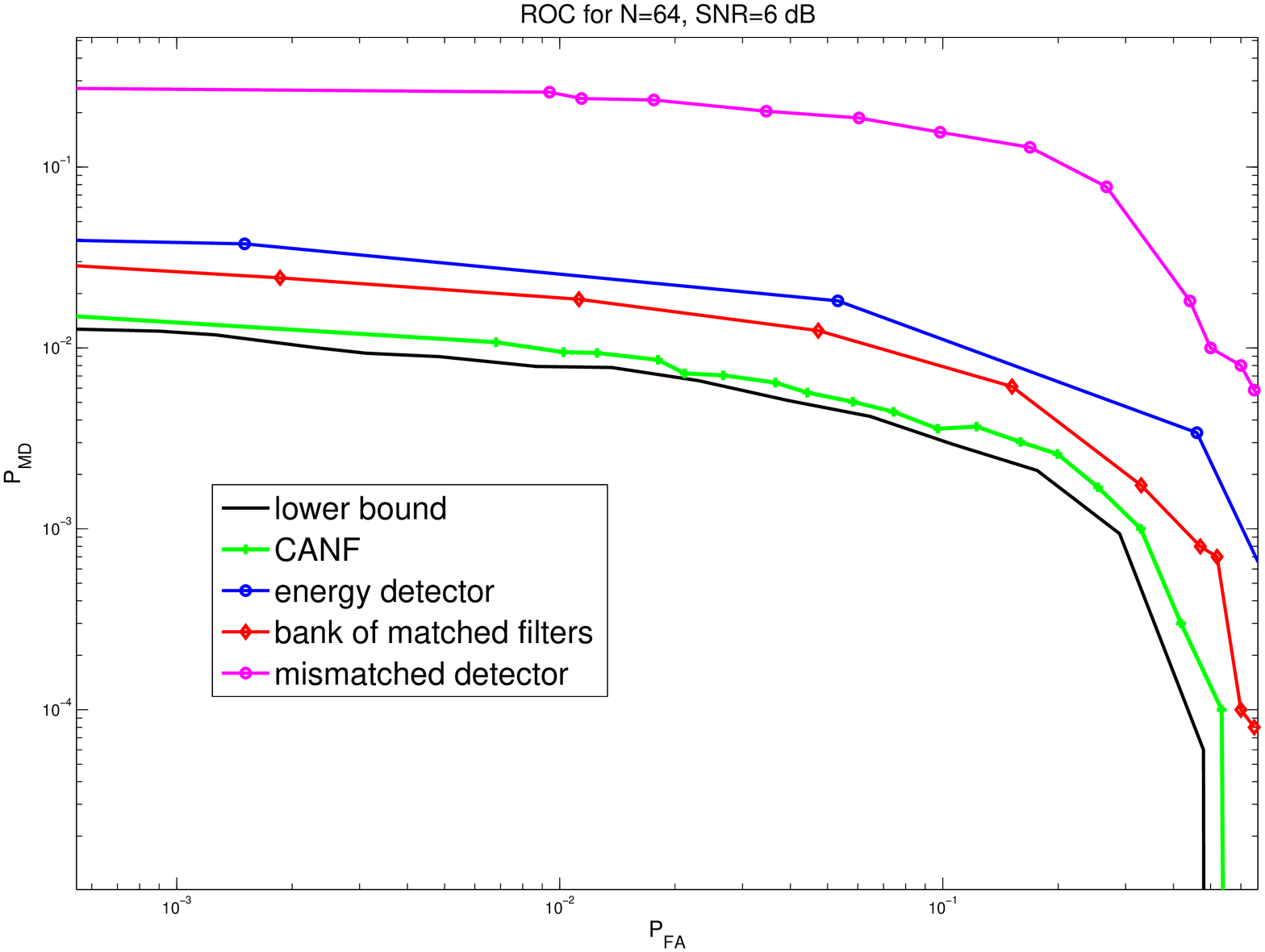}        
        \caption{Comparison of $P_{MD}$ vs. $P_{FA}$ for the proposed algorithm, for $N=64$ and SNR $= \left[0,3,6\right]$dB}
    \label{fig:sensing_N_64}
\end{figure}

\begin{figure}
    \centering
        \epsfysize=7cm
        \epsfxsize=15cm
        \epsffile{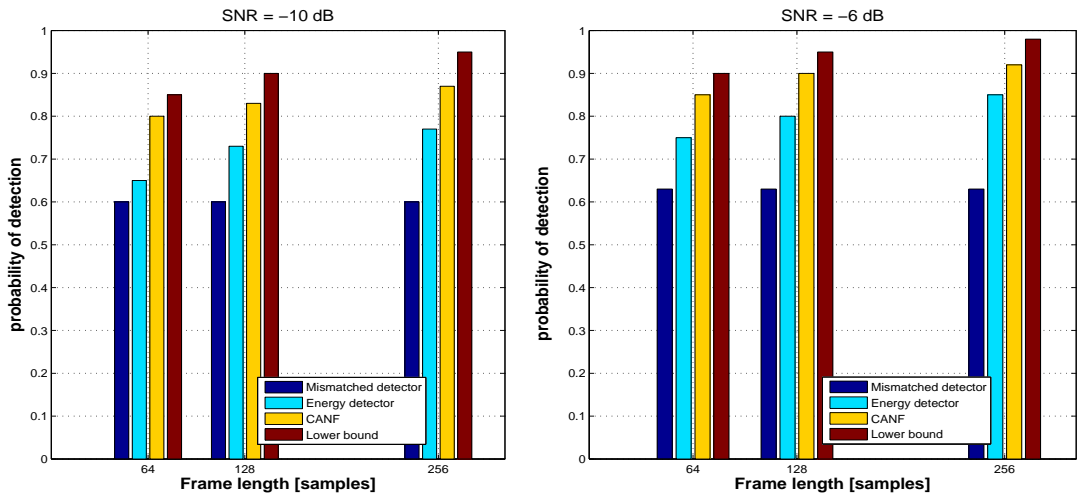}
        \caption{Detection probabilities for various frame lengths ($N= \left\{64, 128, 256\right\}$) and $P_{FA}=0.1$}
    \label{fig:sensing_N_128}
\end{figure}

Next, we compare the proposed algorithm (labeled as CANF) with other detection schemes: the energy detector; the ``close to optimal" solution based on a bank of matched filters as per (\ref{LR_blind_approximate}), with $K=20$ (this makes the computational complexity of this algorithm roughly $4$ times the complexity of the CANF algorithm); a mismatched detector which makes the realised frequency is the nominal one; and as a lower bound we use a detector that has a full knowledge of the realised frequency;
The Receiver Operating Characteristics (ROC) results are presented in Fig. \ref{fig:sensing_N_64}, for various SNR values.
As seen from the results, the proposed algorithm performs much better than the energy and the mismatched detectors. The bank of matched filters provides poorer performance than the CANF detector. It would take $K\approx40$ to achieve similar performance as the CANF detector, which makes its computational complexity around $8$ times more for the same performance characteristics. 
The mismatched filter performs very poorly, which demonstrates how important it is to take into account frequency offsets in the design of the system. As the results depict, the CANF performs close to the lower bound.\\
Next, we fixed $P_{FA}$ to $0.1$ and obtained $P_D$ for various frame lengths and different SNRs. The simulation results are presented in Fig. \ref{fig:sensing_N_128}. We observe that the mismatched detector performs poorly and that increasing the frame length does not improve its performance. We also note that the proposed algorithm performs very close to the lower bound.
\section{Conclusions and Future Work}
In this paper, we proposed a low complexity algorithm for spectrum sensing over fading channels with frequency offsets.
The scheme was based on an adaptive notch filter to perform a low complexity frequency estimation followed by a single matched filter. Simulation results show that the comparable performance to the ``close to optimal" scheme can be obtained with only a fraction of the algorithmic complexity. Future research will include the scenario of collaboration of multiple sensors, and dynamic evolution of the frequency offset.
   
\bibliographystyle{IEEEtran}
\bibliography{../../../references}

\begin{thebibliography}{10}
\providecommand{\url}[1]{#1}
\csname url@samestyle\endcsname
\providecommand{\newblock}{\relax}
\providecommand{\bibinfo}[2]{#2}
\providecommand{\BIBentrySTDinterwordspacing}{\spaceskip=0pt\relax}
\providecommand{\BIBentryALTinterwordstretchfactor}{4}
\providecommand{\BIBentryALTinterwordspacing}{\spaceskip=\fontdimen2\font plus
\BIBentryALTinterwordstretchfactor\fontdimen3\font minus
  \fontdimen4\font\relax}
\providecommand{\BIBforeignlanguage}[2]{{%
\expandafter\ifx\csname l@#1\endcsname\relax
\typeout{** WARNING: IEEEtran.bst: No hyphenation pattern has been}%
\typeout{** loaded for the language `#1'. Using the pattern for}%
\typeout{** the default language instead.}%
\else
\language=\csname l@#1\endcsname
\fi
#2}}
\providecommand{\BIBdecl}{\relax}
\BIBdecl

\bibitem{haykin2005cognitive}
S.~Haykin, ``{Cognitive radio: brain-empowered wireless communications},''
  \emph{IEEE journal on selected areas in communications}, vol.~23, no.~2, pp.
  201--220, 2005.

\bibitem{IEEE802_22}
``{IEEE P802.22/D0.5, Draft Standard for Wireless Regional Area Networks Part
  22: Cognitive Wireless RAN Medium Access Control (MAC) and Physical Layer
  (PHY) specifications: Policies and procedures for operation in the TV
  Bands},'' 2008.

\bibitem{ghasemi2008spectrum}
A.~Ghasemi and E.~Sousa, ``{Spectrum sensing in cognitive radio networks:
  requirements, challenges and design trade-offs},'' \emph{Communications
  Magazine, IEEE}, vol.~46, no.~4, pp. 32--39, 2008.

\bibitem{akyildiz2008survey}
I.~Akyildiz, W.~Lee, M.~Vuran, and S.~Mohanty, ``{A survey on spectrum
  management in cognitive radio networks},'' \emph{IEEE Communications
  Magazine}, vol.~46, no.~4, p.~40, 2008.

\bibitem{yucek2009survey}
T.~Yucek and H.~Arslan, ``{A survey of spectrum sensing algorithms for
  cognitive radio applications},'' \emph{Communications Surveys \& Tutorials,
  IEEE}, vol.~11, no.~1, pp. 116--130, 2009.

\bibitem{vantrees1968dea}
H.~Van~Trees, \emph{{Detection, estimation, and modulation theory.. part 1,.
  detection, estimation, and linear modulation theory}}.\hskip 1em plus 0.5em
  minus 0.4em\relax Wiley New York, 1968.

\bibitem{li2008quickest}
H.~Li, C.~Li, and H.~Dai, ``{Quickest spectrum sensing in cognitive radio},''
  in \emph{Information Sciences and Systems, 2008. CISS 2008. 42nd Annual
  Conference on}.\hskip 1em plus 0.5em minus 0.4em\relax Ieee, 2008, pp.
  203--208.

\bibitem{zheng2010spectrum}
K.~Zheng, H.~Li, S.~Djouadi, and J.~Wang, ``{Spectrum sensing in low SNR regime
  via stochastic resonance},'' in \emph{Information Sciences and Systems
  (CISS), 2010}.\hskip 1em plus 0.5em minus 0.4em\relax IEEE, 2010, pp. 1--5.

\bibitem{li2009adaptive}
C.~Li, H.~Dai, and H.~Li, ``{Adaptive quickest change detection with unknown
  parameter},'' \emph{Proceedings of the 2009 IEEE International Conference on
  Acoustics, Speech and Signal Processing-Volume 00}, pp. 3241--3244, 2009.

\bibitem{lyapunov1994general}
A.~Lyapunov and J.~Walker, ``{The general problem of the stability of
  motion},'' \emph{Journal of Applied Mechanics}, vol.~61, p. 226, 1994.

\bibitem{stoica1988performance}
P.~Stoica and A.~Nehorai, ``{Performance analysis of an adaptive notch filter
  with constrained poles and zeros},'' \emph{IEEE Transactions on Acoustics
  Speech and Signal Processing}, vol.~36, no.~6, pp. 911--919, 1988.

\end{thebibliography}

\newpage

\end{document}